\documentclass[a4paper,journal,transmag]{IEEEtran}

\ifCLASSINFOpdf
\else
\fi
\usepackage[top=30mm, bottom=25mm, left=20mm, right=20mm]{geometry}
\usepackage[cmex10]{amsmath}
\usepackage{framed}
\usepackage[pagebackref=false,colorlinks,linkcolor=black,citecolor=black]{hyperref}
\usepackage{array}
\usepackage{booktabs}
\usepackage{placeins}
\newcolumntype{L}[1]{>{\raggedright\let\newline\\\arraybackslash\hspace{0pt}}m{#1}}
\newcolumntype{C}[1]{>{\centering\let\newline\\\arraybackslash\hspace{0pt}}m{#1}}
\newcolumntype{R}[1]{>{\raggedleft\let\newline\\\arraybackslash\hspace{0pt}}m{#1}}

\ifCLASSINFOpdf
\usepackage[pdftex]{graphicx}
\else
\usepackage[dvips]{graphicx}
\fi
\usepackage{subfig}

\usepackage{algorithm}
\usepackage{algorithmic}

\renewenvironment{abstract}
{\small
	\list{}{%
		\setlength{\leftmargin}{10mm}
		\setlength{\rightmargin}{\leftmargin}%
	}%
	\item\relax}
{\endlist}

\begin{document}
\title{Pre-Hospital Management of Acute Myocardial Infarction Using Tele-Electrocardiography System}

\author{\IEEEauthorblockN{Masoud Elhami Asl\IEEEauthorrefmark{1,*},
Masoomeh Rahimpour\IEEEauthorrefmark{1},
Mahmoud Reza Merati\IEEEauthorrefmark{1}, \\
Amir Hossein Panahi\IEEEauthorrefmark{1,2}, and
Kamran Gholami\IEEEauthorrefmark{1}
}

\IEEEauthorblockA{\IEEEauthorrefmark{1} \footnotesize{Pooyandegan Rah Saadat Co., Tehran, Iran.}}
\IEEEauthorblockA{\IEEEauthorrefmark{2} \footnotesize{IRAN Telecom Research Center,	Tehran, Iran.}}
}

\IEEEtitleabstractindextext{%
\begin{abstract}
\normalfont{\textit{Abstract}--- A comprehensive survey revealed that many patients of Acute Myocardial Infarction reach to the hospital so late to deliver an effective treatment. This leads to poor treatment outcome which increases the mortality rates. Multiple reasons such as lack of diagnostic facilities in the rural health care centers and absence of cardiologists in Emergency Rooms inhibit the timely treatment. In this study, we aimed to develop an effective system to pre-hospital management of Acute Myocardial Infarction. The effectiveness of early thrombolysis in Myocardial Infarction is well established, particularly with regard to its positive effect on reducing the mortality rates. Our proposed system by using tele-electrocardiography technology enables cardiologist to analyze the patient's signal far away from a medical center. This system transmits the collected ECG signal as well some vital signs to the medical center by using internet network. In order to review and analyze the recorded data, an ECG Viewer software is added to this system. The automatic measurement algorithm and some additional tools embedded in the ECG Viewer provide a convenient platform for ECG signal analysis by the cardiologist in the medical center. The designed system has been tested as a prototype in Tehran Emergency Service Center. The evaluation of the proposed system shows acceptable results in recording, transmission, and accurate analysis of ECG signals to early diagnosis of Acute Myocardial Infarction.}
\end{abstract}
}
\maketitle
\IEEEdisplaynontitleabstractindextext
\IEEEpeerreviewmaketitle

\section{Introduction}
Acute Myocardial Infarction (MI) is a major cause of death and disability throughout the world. Approximately 15 million patients per year present to the Emergency Room (ER) with chest pain or other symptoms suggestive of MI~\cite{bassand2007guidelines}. Rapid identification of MI before the golden hour is crucial for the initiation of effective evidence-based medical treatment and management~\cite{thygesen2007universal}. MI occurs when a narrowing in the arteries or a sudden blockage from a blood clot cuts off the nutrients and oxygen supply to the heart muscle. The Golden Hour is a critical time because the heart muscles start to die within 80-90 minutes after it stops getting blood, and within six hours almost all the affected parts of the heart could be irreversibly damaged. On the other hand, only 1/6 of people coming to ER have stroke and the others have some minor inconvenience~\cite{apollolife}. So, the much faster and more effective diagnosis of the heart attack, the lesser would be the damage to the heart, the more chance of recovery the stroke patient and the less time and cost of diagnosis. Electrocardiography (ECG) is among the most useful tools for diagnostic cornerstones and complement the clinical assessment.

The aim of this study is to develop novel approach to early diagnosis of heart attack using emerging tele-medicine technology. Tele-medicine is the use of electronic communications and information technologies to provide clinical services when participants are at different locations. It eliminates distance barriers and improves the access to medical services that would often not be consistently available in distant rural communities.

\begin{figure*}
	\centering 
	\includegraphics[width=12cm]{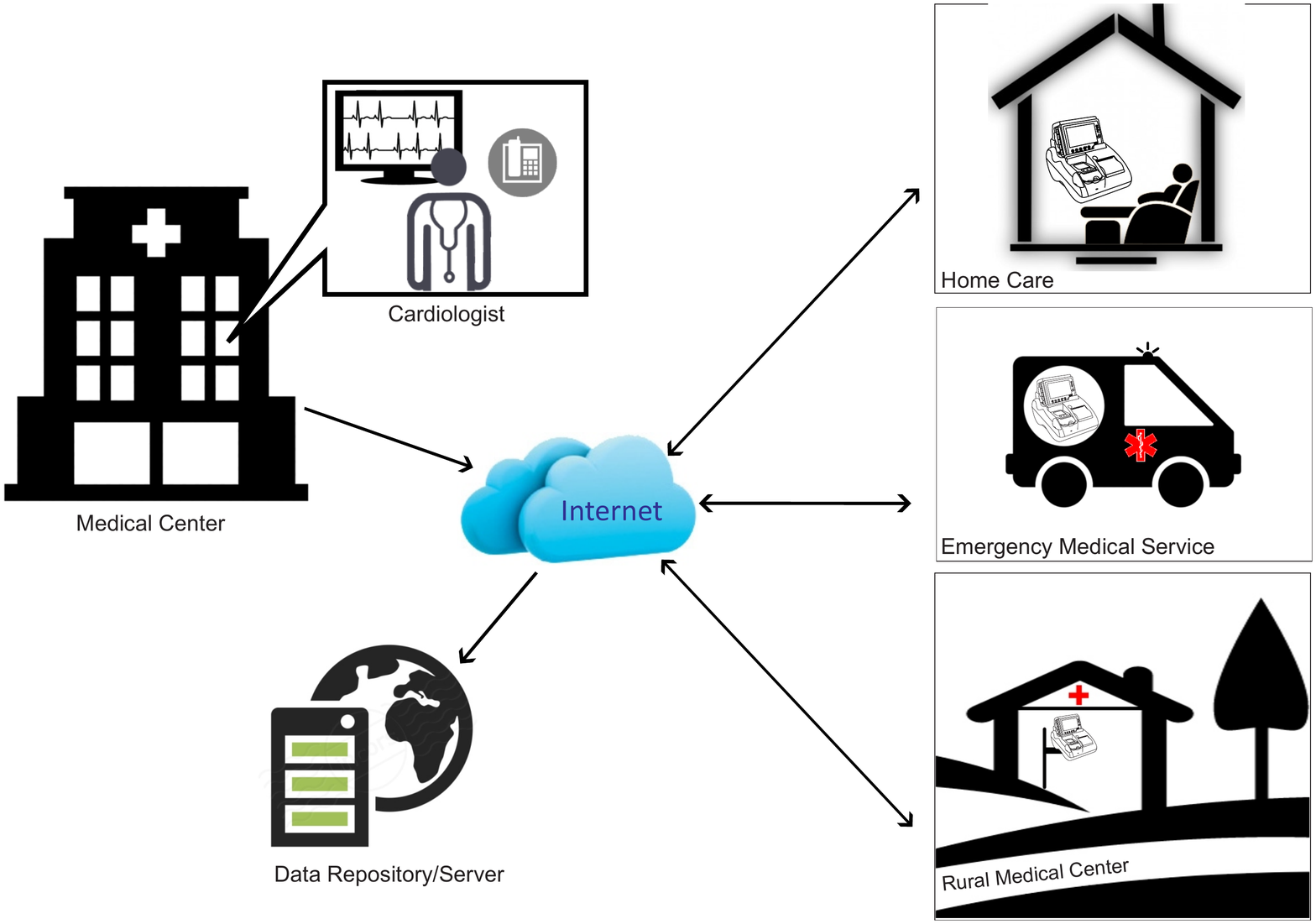}
	\caption{GENERAL BLOCK DIAGRAM OF DEVELOPED SYSTEM.}
	\label{fig:Tele}
\end{figure*} 

The developed system is equipped with both ADSL and mobile data-network infrastructure (3G/4G) to transfer the collected ECG signals and some vital signs including HR and NIBP to the medical center. This system also contains an enhanced ECG viewer software which provides a proper client environment for the storage, analysis, annotating, and reviewing the 12-lead ECG signals. This software includes an automated ECG measurement algorithm and other additional tools that conduct to accurate manual measuring. These facilities make the developed system a powerful tool leading to quick and efficient diagnosis of MI. Moreover, the system can be used in other applications such as rural tele-medical centers and tele-homecare. For populations in distant city hospitals, it is increasingly feasible to provide rural clinics with inexpensive medical instruments such as electrocardiographs that transmit digital ECGs to specialized medical centers and clinics to get the diagnostic information. This system extends the reach of diagnosticians to remote areas within the shortest possible time~\cite{clifford2012signal}. The next sections present a brief description of proposed approach. 


\section{Method and Materials}
Tele-medicine involves the integration of information, telecommunications, human-machine and healt-hcare technologies. One of the aspects of tele-cardiology known as tele-electrocardiography deploys electrocardiography systems to transmit ECG signals and vital signs over internet networks. Tele-cardiology applications can be categorized as pre-hospital, in-hospital and post-hospital. The major purpose of 12-lead electrocardiographic diagnosis is the early detection of MI and to transmit the diagnostic information to the receiving emergency physician before the arrival of the patient to the medical center~\cite{ganguly2000software,sorensen2013telecardiologia}. In this study, we presented an integrated system to readily diagnosis of MI by distance monitoring of the patient. The general block diagram of developed system have been shown in Figure~\ref{fig:Tele} and, the performance of each part have been presented in the following.

\subsection{ECG Signal Recording}
The accurate acquisition of ECG signals is an important step to obtain the reliable diagnostic information. In the introduced system, this step has been done based on the requirements recommended in IEC 60601-2-25 standard~\cite{standard}. The collected data include 12-Lead ECG signals with the sampling frequency of 500 Hz, with the bandwidth of 0.5-150 Hz and 24 bit resolution. Considering the concept of tele-electrocardiography, we modified the sampling rate and the bandwidth of collected data according to IEC 60601-2-25 standard to add the diagnostic capability to the monitoring system. These capabilities include the measurement of global duration, intervals and heart axis.
 
\begin{figure*}
	\centering 
	\includegraphics[width=15cm]{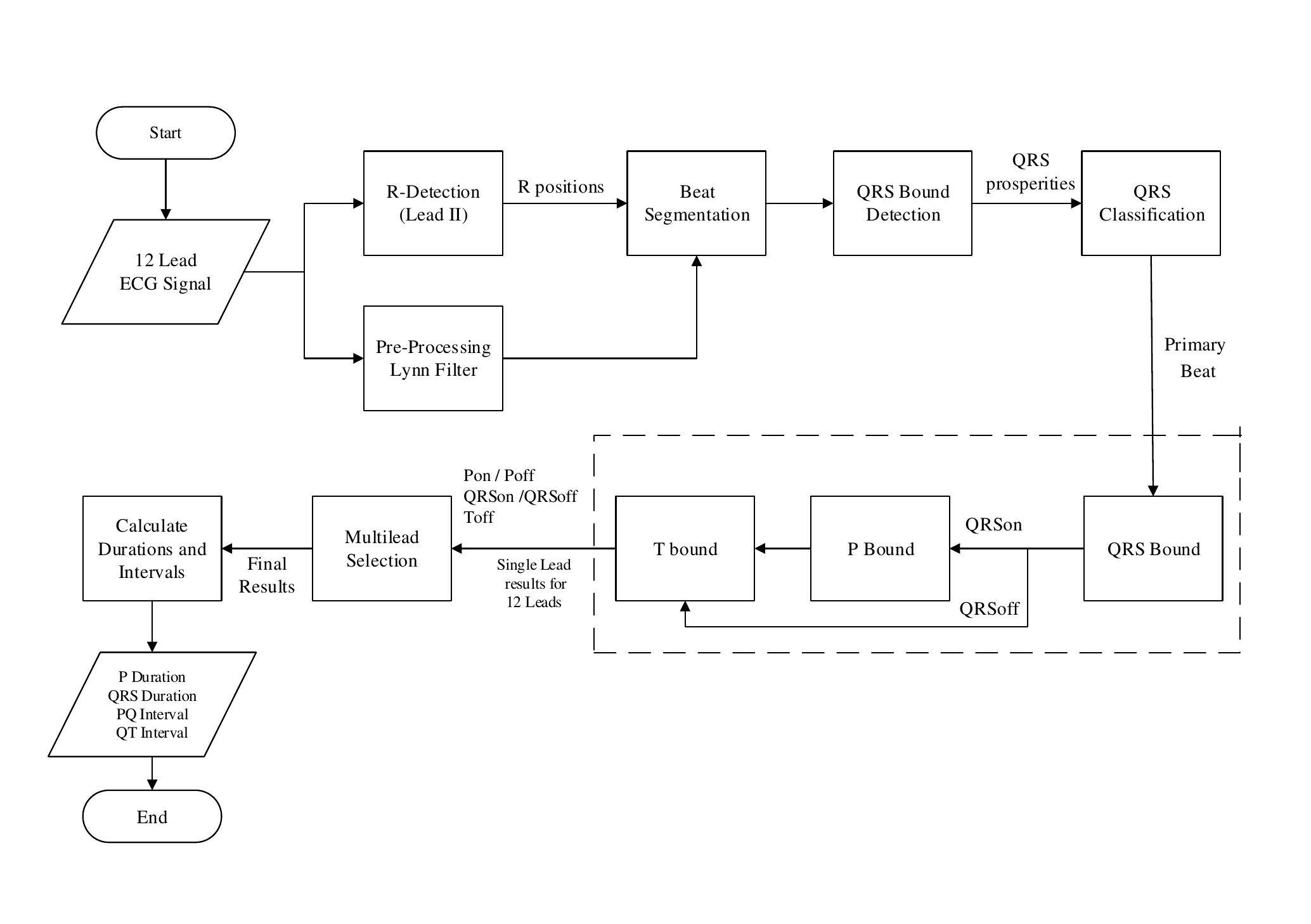}
	\caption{GENERAL BLOCK DIAGRAM OF MEASUREMENT ALGORITHM.}
	\label{fig:Block}
\end{figure*}

\subsection{Data Transmission}
The ECG signals collected in the accident site, home or clinic will be sent to the medical call center using ADSL or mobile data-network infrastructure (3G/4G) in a tag-value format based on HTTP protocol. Data integrity and connection robustness will be guaranteed by TCP/IP protocol. This kind of communication provides a direct contact with cardiologist to get an accurate diagnosis. During the data transmission, the patient demographics will not be transmitted to protect the confidentiality of patients’ identity. Furthermore, to make the data traceable the ID of electrocardiography system will be linked to the ambulance ID along with the recording time and date and the whole data will be sent to the data repository as a single batch. 

\subsection{Medical Center}
After recording the data, ECG signals must be transmitted to the data repository. As a part of tele-electrocardiography system, we need a software to display and analyze the recorded signal. For this purpose, SAADAT ECG Viewer has been designed as an auxiliary tool to help physicians to review and analyze the ECG signals. By using internet connection and ECG Viewer, the cardiologist can download transmitted signals and interpret them easily. In order to facilitate the process of diagnosis, viewer has been equipped with some additional tools such as filters and automated measurement algorithm. In the following, the important tools used in the medical center have been presented.

\subsubsection{SAADAT ECG Viewer}
The 12-Lead ECG signals used for tele-monitoring applications make patient diagnostic information more readily available for both clinician and remote consulting physician. With automated ECG measurements and flexible on-screen reporting functions, the digital ECG enables clinicians to spend less time documenting and more time collaborating with the physician reviewing the results.

SAADAT ECG Viewer has been designed in order to display and analyze the recorded signals in PC by the physicians. Connecting to the network, viewer downloads the ECG signals from data server and displays them to the physicians. In addition, it provides a professional, safe and user friendly client environment for reviewing, analyzing and interpretation of collected ECG signals. Furthermore, this software uses an efficient algorithm for automated measurement of fiducial points in ECG signal which helps physicians interpret ECG signals more quickly and accurately.  

\textit{Filters: }  
In order to improve the signal quality, different kind of filters have been used in the viewer including high-pass and low-pass filters, notch filter, EMG filter, and baseline wandering removal filter. The application of notch filter on ECG signal is to eliminate 50/60 Hz power-line interference. In addition, a high-pass filter with cutoff frequency of 0.8 Hz is defined to remove signal's DC baseline. In order to remove motion artifact noise, a low-pass filter with cutoff frequency of 150 Hz is designed. Since muscle artifact has been a major problem in ECG monitoring and electrocardiography systems, a dynamic filter with variable smoothing effect in different parts of the signal is used in order to suppress the EMG noise~\cite{hashemi2015dynamic}. The wide band frequency spectrum of muscle noise and its overlap with ECG signal make it impossible to suppress it with regular filters without distortion of ECG signal. This filter suppresses EMG noise outside the QRS region while applying much less smoothing to the QRS complex. All these filters make ECG signal more clear and legible for physicians by removing disruptive information.
\FloatBarrier

\begin{figure*}
	\centering 
	\includegraphics[width=15cm]{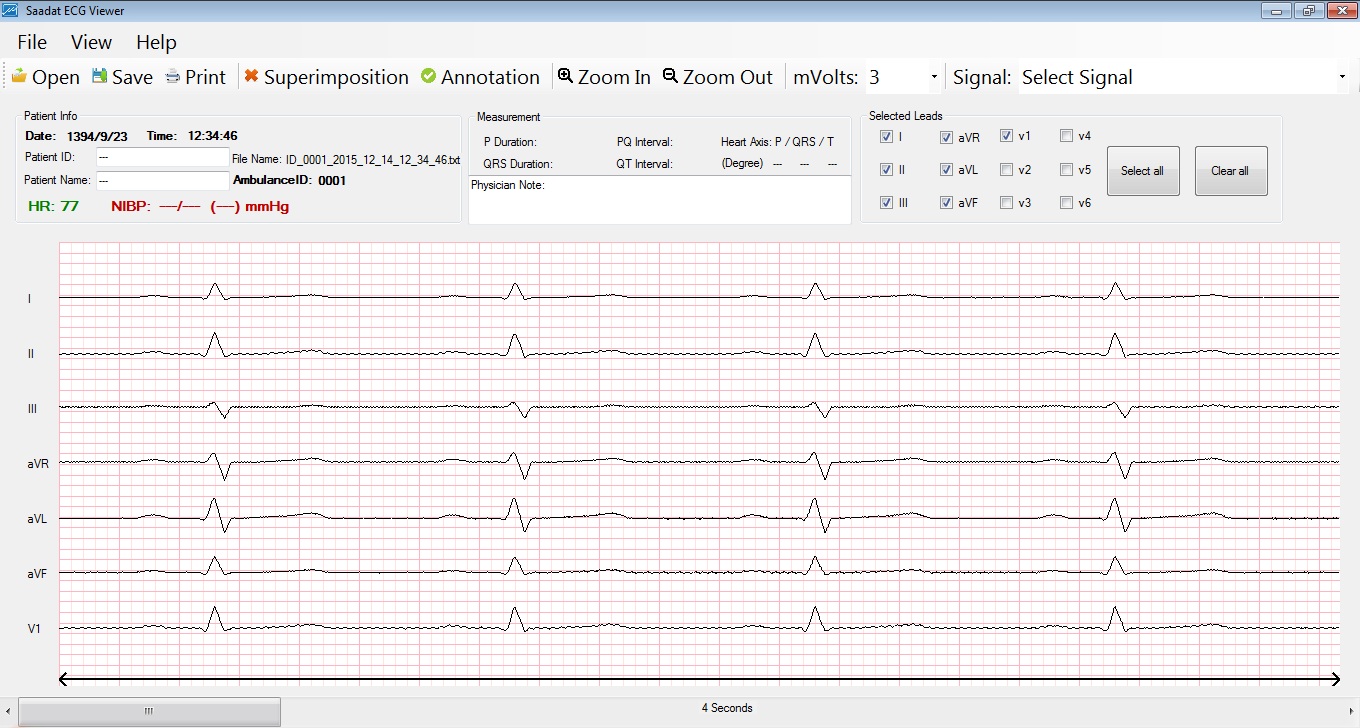}
	\caption{SAADAT ECG VIEWER.}
	\label{fig:Viewer}
\end{figure*}

\textit{Measurement Algorithm:}
Even though studies showed that the accuracy of computerized ECG measurement and interpretation is less than cardiologists’ decision, using such algorithms as an auxiliary tool has a remarkable effect on improving their diagnosis. It also helps primary care physicians interpret the ECG signals when cardiologists are not accessible. Therefore, the automated measurement algorithm has been added to SAADAT ECG Viewer to calculate important parameters of ECG signals.
Proposed measurement algorithm is based on analyzing 10-second 12-lead ECG signals. To remove different artifacts and improve the quality of signal, a pre-processing step has been applied to the input signal. The next step of measurement algorithm is R wave detection which is necessary to beat segmentation and is accomplished based on Pan-Tompkins algorithm~\cite{pan1985real} applying on the signal of lead II; This is a well-known algorithm performing based on digital analysis of the slope, amplitude and width which has been used in most of the recent successful methods for QRS detection in automated ECG signal analysis~\cite{lin2010p}~\cite{rahimpour2016p}. The accurate location of R peaks on the other leads is determined based on windowing technique around the approximated peak points called $PK_i$. The general block diagram of measurement algorithm has been shown in Figure~\ref{fig:Block}.
According to the peak point detected as an approximated R wave, each beat has been segmented and analyzed to estimate the onset and offset points of each wave and its amplitude. Extracting main features of QRS complex including QRS duration, R-R interval and R wave peak to peak amplitude is the next step which identifies the class to which each complex belongs. This classification logic has to allow for a single normally conducted beat and is similar to method used in Kenz electrocardiography system. Three classes are considered in order to beat classification including "0" type as a normal beat, "1" type as a normal beat with the shorten R-R interval which has laid before the abnormal beat, and "2" type as an abnormal ECG beat. After beat classification, only one beat will be selected as a primary beat to be analyzed according to dominant method described in~\cite{Kenz2009}. The next processing step is QRS typing. Generally, 6 different types have been considered for QRS complex; these are including RSR', QR, QRS, RS, R and QS. Identifying the type of QRS complex is done according to the initial estimation of peak points on each complex known as $PK_i$; these points are the candidates of Q, R, S and R' waves according to their amplitude and upward or downward slopes. After type declaration of QRS complex, the onset and offset points of QRS wave have been determined within the windows locating according to the R wave location. In order to detect P and T wave boundaries, the information of slope signal has been used. Using moving window for P wave detection has decreased the number of falsely detected points in the case of large PR intervals. In addition, an adaptive thresholding in conjunction with a varied sized window for T wave detection make the algorithm more robust. The length of this window for each wave can be changed according to the wave morphology.

Physiologically, the global duration of P, QRS and T waves are defined by the earliest onset in one lead and the latest offset in any other lead; indeed wave onset and offset do not necessarily appear at the same time in all leads, because the activation propagate differently~\cite{standard}. Considering this issue, we used a Multi-lead algorithm to select the final value of parameters; this method has reduced the influence of possible noisy measurements. Interested reader referred to~\cite{rahimpour2016ecg} for detailed algorithm.

The measurement algorithm has been implemented in Visual Studio C++ 2010. Using CSE Multi-lead database, the performance of this algorithm has been evaluated. CSE (Common Standard for Quantitative Electrocardiography) is a reference database are being used by more than 110 academic and industry research centers in order to assess and improve ECG measurement and interpretation programs. This database contains 125 records each of them lasting 10 seconds with the sampling frequency of 500 Hz~\cite{willems1985common}. 

\begin{figure}
	\centering 
	\subfloat[]{\includegraphics[width=7cm]{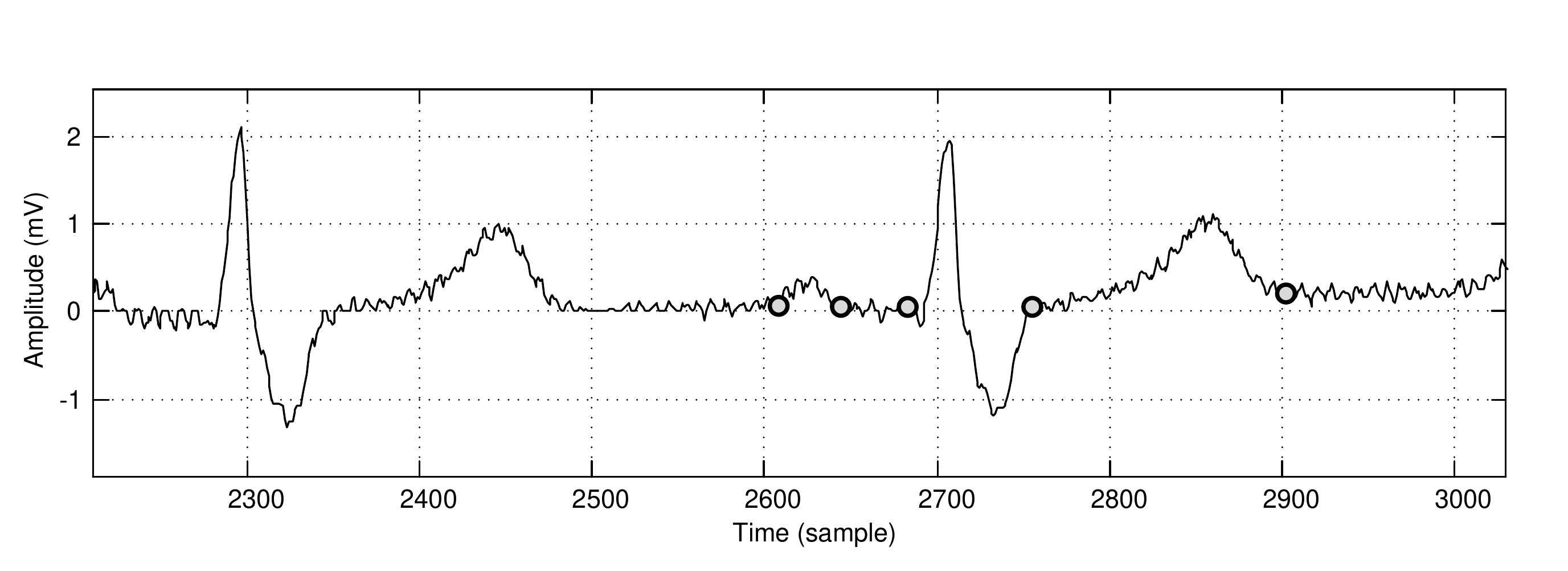}}
	\\
	\subfloat[]{\includegraphics[width=7cm]{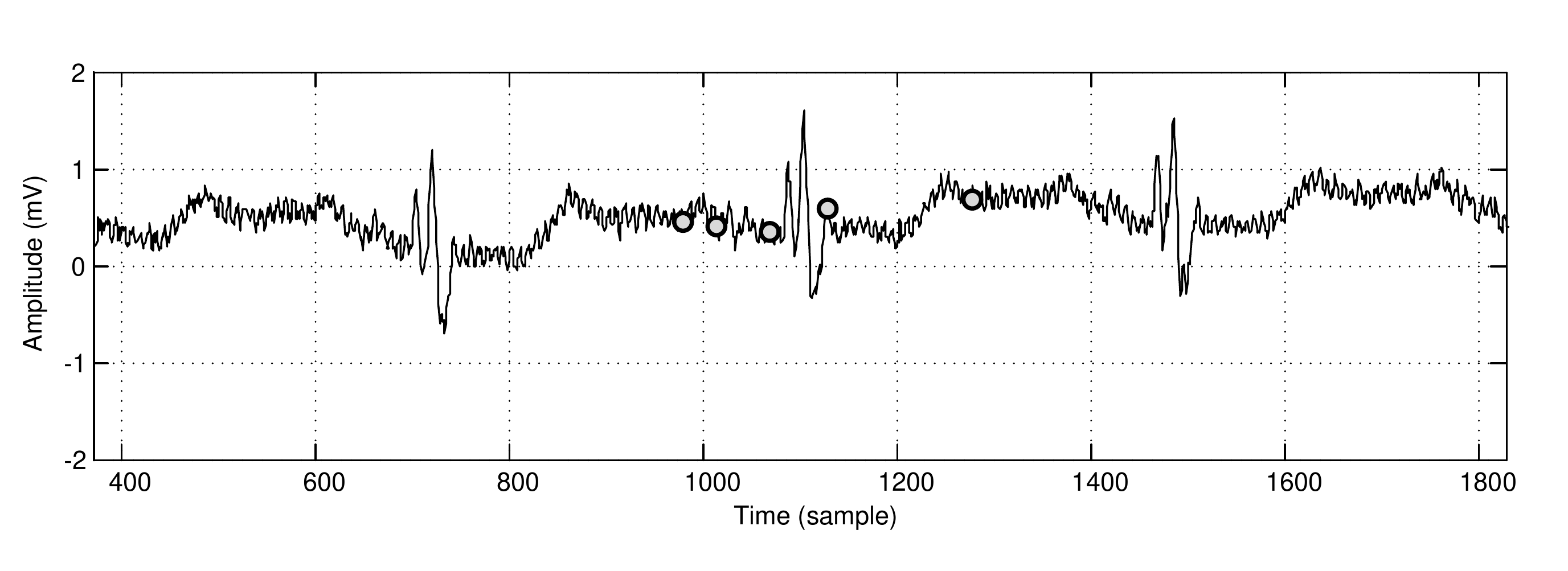}}
	\\
	\subfloat[]{\includegraphics[width=7cm]{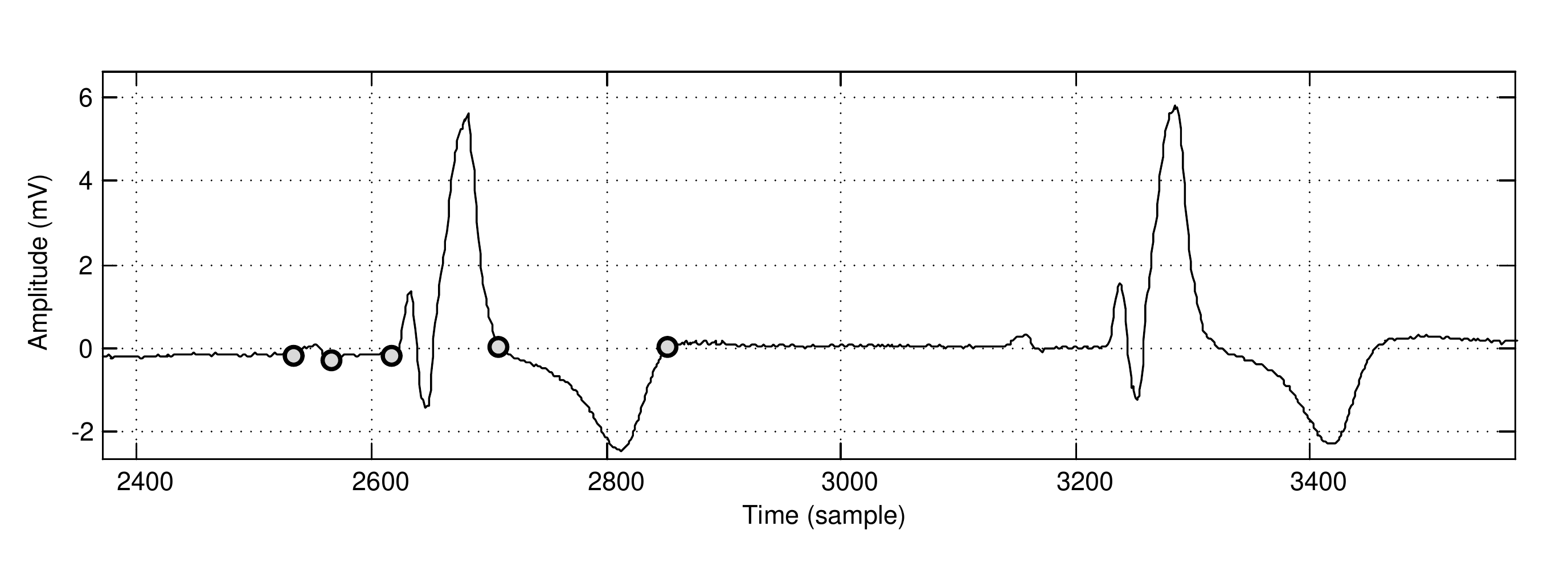}}
	\caption{TYPICAL RESULTS OF PROPOSED ALGORITHMS FOR DIFFERENT ECG SIGNALS.}
	\label{fig:result}
\end{figure}

Using the proposed method, the duration and intervals of ECG signals have been calculated for CSE database. Deviation of these measurements from the mean referee estimates of CSE database are presented in Table~\ref{table:1}; the acceptable limits for these parameters recommended in Table 201.105 of IEC-60601-2-25 standard~\cite{standard} are mentioned in these tables. Figure~\ref{fig:result}, also shows the typical results of the algorithms for different ECG signals.

\begin{table}
	\renewcommand{\arraystretch}{1.4}
	\caption{THE RESULT OF ECGPUWAVE VS PROPOSED METHOD ON CSE DATABASE.}
	\begin{center}
		\begin{tabular}{|@{}C{1.5cm}@{}|@{}C{1cm}@{}|@{}C{1cm}@{}|@{}C{1.2cm}@{}|@{}C{1cm}@{}|@{}C{1cm}@{}|@{}C{1.2cm}@{}|}
			\hline \hline
			{} & \multicolumn{3}{c|}{\textbf{Mean Difference (ms)}} & \multicolumn{3}{c|}{\textbf{Standard Deviation}}\\
			\hline 
			{} & Standard & ECGpuwave & Proposed method & Standard & ECGpuwave & Proposed method \\
			\hline 
			P Duration   & $\pm$10 & 7.35  & 0.70 & 15 & 10.27 & 3.70 \\ 
			QRS Duration & $\pm$10 & 7.77  & 1.40 & 10 & 14.51 & 3.52 \\ 
			PQ Interval  & $\pm$10 & 2.5   & -1.1 & 10 & 10.73 & 3.04 \\ 
			QT Interval  & $\pm$25 & -7.57 & 4.09 & 30 & 35.29 & 5.61 \\ 
			\hline \hline
		\end{tabular}
	\end{center}
	\label{table:1}
\end{table}

\textit{Heart axis:} 
In clinical applications, heart axis provides helpful information for medical diagnosis. The electrical axis of the heart is the mean direction of action potentials traveling through the ventricles during ventricular activation (depolarization). The QRS complex which represents ventricular depolarization is more common to determine the frontal electrical heart axis. Any combination of 2 limb leads can be used to calculate the QRS frontal axis. Using the proposed viewer, not only the heart axis can be calculated automatically, but also they would be calculated manually by choosing each components’ baseline and peak values. Deviation of these measurements from the mean referee estimates of CSE database are presented in Table~\ref{table:2}.

\begin{table}
	\renewcommand{\arraystretch}{1.4}
	\caption{MEAN DIFFERENCE AND STANDARD DEVIATION FOR P, QRS, AND T AXIS ON CSE DATABASE.}
	\begin{center}
		\begin{tabular}{|@{}C{1.5cm}@{}|@{}C{1cm}@{}|@{}C{1.2cm}@{}|@{}C{1cm}@{}|@{}C{1cm}@{}|@{}C{1.2cm}@{}|@{}C{1cm}@{}|}
			\hline \hline
			{} & \multicolumn{3}{c|}{\textbf{\scriptsize{Mean Difference (in degree)}}} & \multicolumn{3}{c|}{\textbf{\scriptsize{Standard Deviation}}}\\
			\hline 
			{} & P axis & QRS axis & T axis & P axis & QRS axis & T axis \\
			\hline 
			CSE DB   & 2.99 & 0.59 & 1.05 & 21.54 & 15.44 & 19.18 \\ 
			\hline \hline
		\end{tabular}
	\end{center}
	\label{table:2}
\end{table}

Computerized ECG analysis appeared to benefit primary care physicians most by providing a backup opinion; this second opinion was also of use to cardiologist. Additional long-term benefits that may be derived from computer systems include improvement of physician interpretation ability, reduction in interpretation time, and standardization of electrocardiographic nomenclature and criteria. Studies shown that primary care physicians altered their initial interpretations in 45 percent after being provided with computer readings of each ECG signal. In addition, experts altered their initial interpretations in 39 percent tracing~\cite{grauer1989potential}.

\FloatBarrier

\textit{Additional tools:} 
Even though the viewer can estimate the parameters of ECG signal automatically, it provides some tools such a ruler and zoom option for physician in order to accurate measurement of parameters manually. Sometime the existence of noise and some kind of unpredictable ECG morphologies may lead to partially unreliable results; in this case these additional tools can help physician. A view of SAADAT ECG Viewer has been shown in Figure~\ref{fig:Viewer}.

\subsubsection{Treatment Measures}
Analyzing the collected ECG signal, cardiologist can diagnosis the type of cardiovascular diseases (CVD) especially MI and call primary care physician to start treatment quickly which drastically reduce diagnosis time. Being able to start treatment expeditiously would be very helpful in accelerating the treatment process. 

\section{Discussion}
Rapid diagnosis of MI is undoubtedly a challenging issue, and in most cases the recommended golden hour of 90 minutes has rarely been attained~\cite{krumholz2008campaign}. The features of tele-electrocardiography are potentially suitable for the development of improved diagnostic systems. Taking the advantages of this technology, we presented a new approach to speed up the diagnostic process of MI and increase the chance of recovery for ischemic stroke patients. The designed system has been tested as a prototype in Tehran Emergency Service Center. For this implementation, a reliable platform is required to provide the basic connectivity infrastructure and a communication protocol for safe transmission of patient’s data. We have chosen both ADSL and mobile data network infrastructure (3G/4G) as the existing data connection networks. To facilitate the analysis of received data rapidly and accurately, we developed an ECG viewer software which provides a suitable environment to automated and manual measurement of the parameters of ECG signals.  To be properly addressed the lake of treatment time, we equipped the developed system with a mobile cell-phone module that enables the cardiologists in medical center to make contact with primary care physician and therefore deliver time effective treatment in accident site before arriving to the hospital. 

Clinical evaluations indicates that presented system and automated ECG analysis are time efficient, reliable and substantially cheaper in cost in comparison to conventional referral systems. It enables a more widespread access to rapid reperfusion thereby; therefore, reducing treatment delay would reduce the morbidity and mortality. 

Considering the effectiveness of proposed system, there are some limitations which should be tackled appropriately. One of them is the internet bandwidth limitations that causes the low speed connection and inability to quick transmission of data to the medical center. One of the solutions to overcome to this issue is using compression techniques to faster sending over low-speed internet connection. Another approach in future studies is considering the privacy issue such as mutual authentication and data redirection. 

The other limitation of using only electrocardiography system is to the possibility of getting noisy ECG signals which make the diagnostic information less reliable. Although the designed filters enhance signal components that are important for analyzing ECG signals, using some additional parameters which help physicians for diagnose of MI will be helpful. Considering the remarkable variations of four blood factors including Myoglobin, Creatine Kinase, Troponin, and H-FABP immediately after MI, an effective solution is to design a module to accurate measuring of them. In the future studies, we are going to examine the effect of immediate measuring and monitoring these diagnostic factors along with electrocardiography records to provide prognostic information in patients with MI. 

\section{Conclusion}
In this study, we developed a tele-electrocardiography system to early diagnosis of MI. Clinical tests performed to evaluate this system show acceptable results for early diagnosis of MI. However, an evaluation strategy based on functional, clinical and economic factors are under consideration. In the future, tele-electrocardiography is expected to further impact on providing better care for the patient with cardiovascular disease.

\begin{figure}
	\centering 
	\subfloat[]{\includegraphics[width=4.8cm]{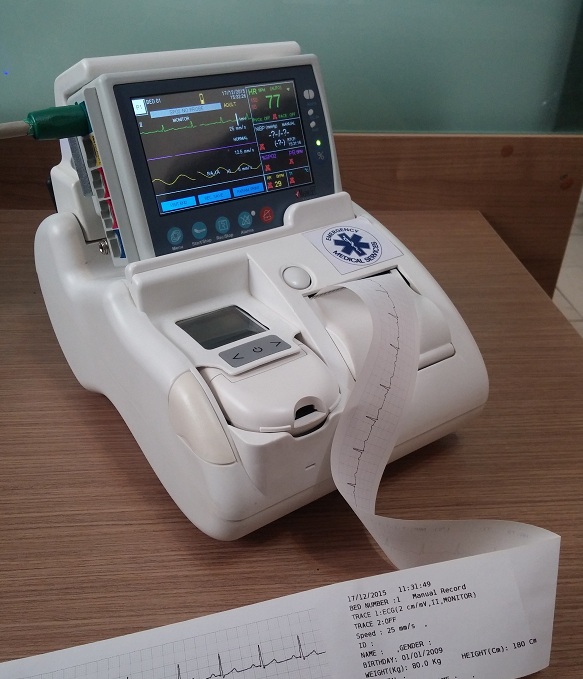}}
	\\	
	\subfloat[]{\includegraphics[width=7cm]{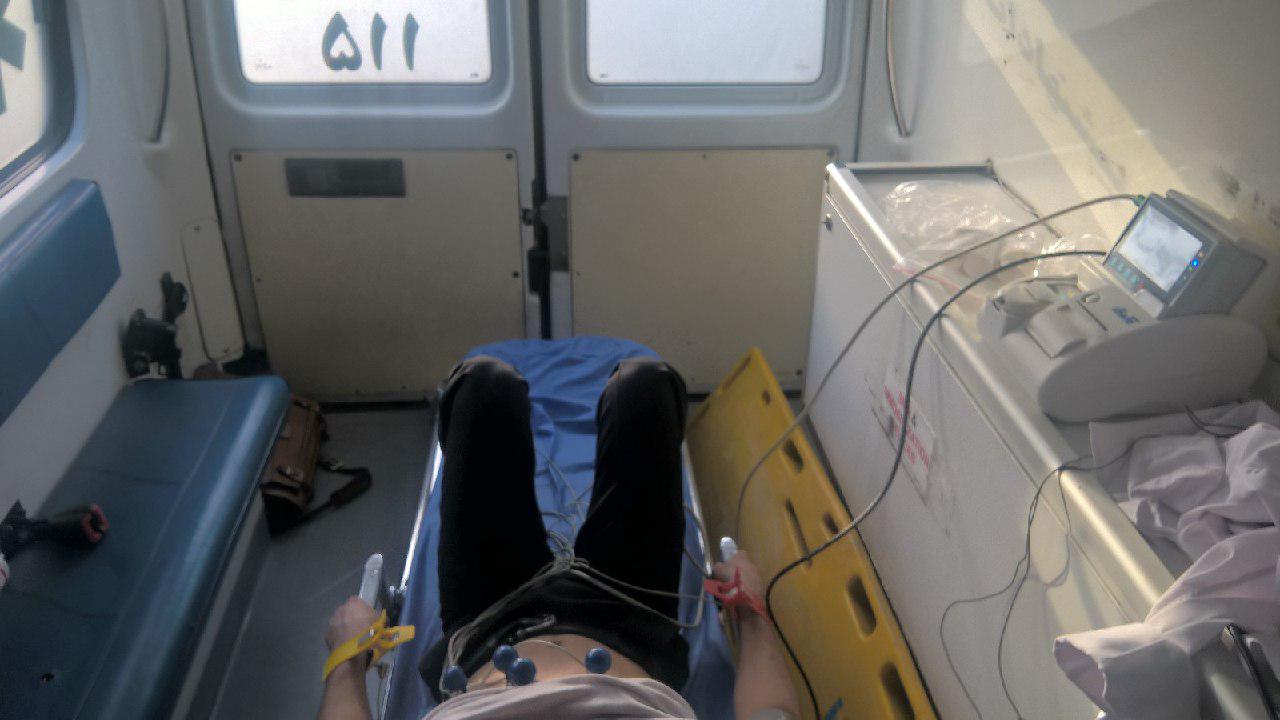}}
	\caption{(a) THE SCHEMATIC OF 12-LEAD ECG SIGNAL RECORDING AND  TRANSMISSION SYSTEM, (b) TEST PROCESS OF DEVELOPED SYSTEM IN AN AMBULANCE.}
	\label{fig:Aria}
\end{figure}

\bibliographystyle{IEEEtran}
\bibliography{MyReferences.bib}

\end{document}